\documentclass{JFM-FLM_Au}
\usepackage{subcaption}

\lefttitle{L.P. Knudsen, J.O. Wenegrat, J.P. Hilditch, L.N. Thomas}
\righttitle{Journal of Fluid Mechanics}
\normalsize

\title{Parametric Subharmonic Instability in the Ocean Bottom Boundary Layer}

\author{L.P. Knudsen\aff{1}, J.O. Wenegrat\aff{1}, J.P. Hilditch\aff{2} \and L.N. Thomas\aff{2}}

\affiliation{\aff{1} Department of Atmospheric and Oceanic Science, University of Maryland, College Park, MD
\aff{2}Department of Earth System Science, Stanford University, Stanford, CA}

\corresau{J.O. Wenegrat, \email{wenegrat@umd.edu}}

\begin{document}
\maketitle

\begin{abstract}
Internal waves with frequency larger than twice the local minimum allowable wave frequency can be susceptible to parametric subharmonic instability (PSI). This instability draws energy from the wave and provides a mechanism for generating small-scale turbulence and mixing. In the ocean, strongly baroclinic flows at the submesoscale adjust the minimum frequency of internal waves such that it is possible for PSI to occur for locally near-inertial waves. One setting where this may occur is in baroclinic bottom boundary layers along sloping topography, where 
near-bottom interior flows in the sense of Kelvin-wave propagation lead to a reduction of bottom boundary layer Ertel potential vorticity, and consequently lower the minimum frequency sufficiently to allow PSI. Linear stability analysis, and nonlinear simulations, show that PSI grows at a rate determined by the vertical stratification of the bottom boundary layer, and the slope Burger number. Wave shear production is the primary energy source for the instability, with additional contributions from buoyancy production that depend on the slope parameters. A partially compensating loss of energy to geostrophic shear production becomes increasingly important as the flow approaches the marginally stable state. These results suggest PSI as a potential mechanism for generating near-bottom mixing in the ocean.
\end{abstract}

\begin{keywords}
Authors should not enter keywords on the manuscript, as these must be chosen by the author during the online submission process and will then be added during the typesetting process (see \href{https://www.cambridge.org/core/journals/journal-of-fluid-mechanics/information/list-of-keywords}{Keyword PDF} for the full list).  Other classifications will be added at the same time.
\end{keywords}


\section{Introduction}
Internal waves in the ocean generate near-bottom mixing that is understood to play a central role in maintaining the overturning circulation \citep{Ferrari_et_al_2016, polzin_mixing_2022}. The dynamical routes from waves to turbulence are many \citep{mackinnon_climate_2017, buijsman_multiagency_2026}, including well-established mechanisms such as the convergence of wave energy at critical slopes (where reflected wave characteristics parallel the bottom), and direct nonlinear wave-breaking when strong waves encounter steep slopes \citep{wynne-cattanach_observations_2024, whitley_breaking_2025}. At the same time a variety of recent work has highlighted that the bottom boundary layer can support a range of strongly baroclinic submesoscale flows \citep{Wenegrat_et_al_2018, Naveira_Garabato_et_al_2019, Wenegrat_et_al_2020, Qu_et_al_2021}, which are known from the surface ocean to significantly modify wave physics \citep{Whitt_and_Thomas_2013}. Here we show that in some conditions this can lead to parametric subharmonic instability (PSI) in the ocean bottom boundary layer, allowing near-inertial wave energy to be drained into growing submesoscale instabilities and subsequent secondary turbulence---a novel route to mixing along the bottom.


Submesoscale flows in the ocean are defined by Rossby and Richardson numbers of order unity, $\text{Ro}\sim \text{Ri}\sim \mathcal{O}(1)$, highlighting that the submesoscale is a unique dynamical regime that sits between balanced rotationally-dominated flows and smaller-scale three-dimensional turbulence \citep{Thomas_et_al_2008, taylor_submesoscale_2023}. Much of the work to-date on submesoscale processes has been in the context of the surface boundary layer, however observations of benthic boundary currents also indicate flows that are dynamically submesoscale \citep{Madron_Weatherly_1994, STAHR1999, Ruanetal2017}. For example, the North Atlantic Deep Western Boundary Current at the Blake Outer Ridge near the continental shelf was shown to have strong horizontal buoyancy gradients and weak near-bottom stratification in a layer extending approximately 250 m off the bottom \citep{STAHR1999}. This layer was argued to be the result of mixing due to a down-slope Ekman buoyancy transport associated with the along-isobath interior flow in the direction of Kelvin-wave propagation. Similar conditions have been observed in other deep boundary current systems \citep{Madron_Weatherly_1994, Naveira_Garabato_et_al_2019}, suggesting processes which generate turbulent mixing in these conditions may play a particularly important role in watermass transformation along the pathways of the overturning circulation. 

The dynamics of these baroclinic layers along topography are strongly influenced by their signature in Ertel Potential Vorticity (PV). The destratifying downslope flow can be understood as a frictional destruction of PV at the boundary \citep{Benthyusen_and_Thomas_2012}. When the PV is negative in the Northern Hemisphere the BBL is susceptible to symmetric instability (SI), which will quickly bring the boundary layer back to a state with near 0 PV \citep{Allen_Newberger_1998, Wenegrat_et_al_2020}. The flow can then become unstable to baroclinic instability, although the growth rates are strongly suppressed for steep topography \citep{brink_instability_2013, Wenegrat_et_al_2018}. The lowering of boundary layer PV also modifies internal wave physics \citep{Whitt_and_Thomas_2013}, including a change in the minimum allowable wave frequency which is reduced from the local inertial frequency, $f$, whenever the PV is lower than the planetary PV (ie. $fN^2$ where $N^2 = \partial b/\partial z$). Work on the surface boundary layer has shown that the reduction of minimum frequency due to low PV at fronts can allow for the development of PSI of near-inertial waves, a wave-wave interaction that occurs when the minimum permitted frequency of an internal wave is half the local inertial frequency \citep{ThomasTaylor14, Hilditch_Thomas_2023}. PSI transfers energy from the larger-scale parent wave to smaller-scale subharmonic waves at half the internal wave's frequency, driving exponential growth and ultimately a transition to turbulence through secondary shear instabilities \citep{McComas1977, MacKinnonWinters05, ThomasTaylor14}. PSI has not been studied in oceanic BBLs, despite the structural similarities of baroclinic flows near the bottom suggesting PSI may occur.

The goal of this study is to show that near-inertial waves in the baroclinic BBL can undergo PSI. In Section \ref{section:theoryanalysis} an idealized baroclinic BBL flow is introduced, along with the equations of motion for a near-inertial oscillating BBL where cross-slope buoyancy gradients influence the frequency and polarization of the oscillations. Floquet analysis is then applied to determine the regions of parameter space that are PSI unstable, as well as the energetics of the instability growth.  Idealized numerical simulations are introduced in \ref{section:numericalmodel}, and then compared to linear theory in \ref{section:simulationoutput}. Select case studies with different sets of parameters are then compared to explore how the choice of parameters affects the energetics and growth of PSI. A summary of results and discussion of implications is given in section \ref{section:conclusions}.

\section{Theory}\label{section:theoryanalysis}

\subsection{Background Flow}\label{section:backgroundflow}
\subsubsection{Rotation of coordinate system}
\begin{figure}[t!]
    \centering
    \includegraphics[width=0.8\linewidth]{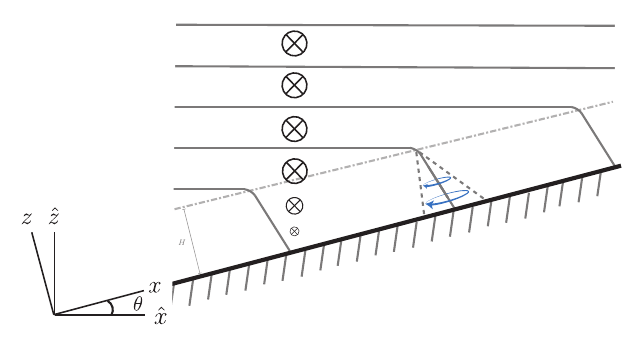}
    \caption{Schematic of the domain and background flow field. Gray lines denote isopycnals which slope downwards over the boundary layer of height $H$ (dash-dot line), associated with a thermal-wind adjustment of the along-isobath flow from the interior velocity to $0$ at the boundary. Near-inertial oscillations (blue arrows) in the boundary layer displace isopycnals (dashed gray lines). The coordinate system, rotated by angle $\theta$ to match the seafloor slope, is indicated at left. }
    \label{fig:slope_coords}
\end{figure}

\qquad We consider the case of a planar sloping bottom, where the flow is invariant in the along-isobath (along-slope) direction. To simply the derivation of the background conditions we formulate a coordinate system rotated by the angle $\theta$ around the $y$-axis such that the across-slope ($x$-axis) and slope-normal ($z$-axis) directions align with the slope as seen in Figure \ref{fig:slope_coords}. The vertical coordinate is defined to be $z=0$ along the bottom. The along-slope direction ($y$-axis) will be invariant to rotation, and the flow is assumed to be uniform in the along-slope direction. The new rotated coordinate system is thus defined to be \citep{Trowbridge_Lentz_1991,Allen_Newberger_1998,Wenegrat_et_al_2018}
\begin{subequations}
    \begin{equation}
        x = \hat{x}+\theta \hat{z}, \text{ and}
    \end{equation}
    \begin{equation}
        z = \hat{z}-\theta \hat{x},
    \end{equation}
\end{subequations}
where the carat notation indicates the unrotated Cartesian coordinates. For simplicity we make the small-angle approximation ($\cos\theta \approx 1$, $\sin\theta \approx \theta$) throughout, consistent with real seafloor slopes \citep{Costelloetal2010}.

\subsubsection{Arrested Ekman Layer}
A BBL with reduced stratification (low PV) is assumed to be in thermal wind balance such that there is a linear geostrophic shear, $\Lambda$, decreasing from the value of the along-isobath interior flow, $V_\infty$, at the top of the boundary layer to 0 at the bottom (Fig. \ref{fig:slope_coords}). This setup, known as the arrested Ekman layer \citep{MacCready_Rhines_1993}, is chosen as a simple representation of a baroclinic BBL amenable to isolating PSI from BBL adjustment dynamics, rather than as an entirely accurate representation of the observed BBL over slopes. Limitations of this are discussed in Section \ref{section:conclusions}. 

The arrested Ekman layer solution can be characterized by several key parameters. The stratification in the boundary layer, $N^2_{BBL}$, can be expressed using a parameter,
\begin{equation}
    \gamma = 1 - \frac{N^2_{BBL}}{N^2_\infty},
\end{equation}
where $N^2_\infty$ is the stratification in the interior such that $\gamma>0$ implies reduced BBL stratification \citep[following][]{Allen_Newberger_1998}. The thermal wind-shear is then given by $\Lambda = N^2_\infty\theta\gamma/f$. Following from this the height of the arrested Ekman layer is given by $H=V_\infty/\Lambda$ \citep{Allen_Newberger_1998}.  This leads to non-dimensional parameters that include the Rossby and Richardson numbers which can be written as \citep{Allen_Newberger_1998},
\begin{subequations}\label{eq:dmnmbrs}
    \begin{equation}
        \text{Ro} = -\frac{\theta \Lambda}{f} = -S^2_\infty \gamma, \text{ and}
    \end{equation}
    \begin{equation}
        \text{Ri} = \frac{N^2_\infty(1-\gamma)}{\Lambda^2} = \dfrac{1-\gamma}{S_\infty^2 \gamma^2},
    \end{equation}
\end{subequations}
and interior slope Burger number,
\begin{equation}
    S^2_\infty = \frac{N^2_\infty\theta^2}{f^2}.
\end{equation}
 For values typical of observed ocean BBLs both $Ro\sim Ri\sim 1$, again emphasizing that these are dynamically submesoscale flows. 

\subsubsection{Near-inertial oscillations in the boundary layer}
Sheared near-inertial oscillations are imposed on the balanced background flow to model an oscillatory arrested Ekman layer and provide the parent wave needed for PSI. The set-up outlined here was chosen to study an Ekman adjustment problem as seen in \cite{Brink_Lintz_2010a,Brink_Lintz_2010b}, where oscillations are generated either as an initial transient or as a response to oscillatory interior flow. More generally we expect similar near-inertial oscillations to be common features generated for instance by time-varying interior flow (including low-frequency variability in deep boundary currents, or mesoscale eddies impinging on topography), along-flow changes in topography (which break balance), or relaxation of localized near-bottom mixing events \citep{legg_internal_2008, whitley_breaking_2025}. The related problem of remotely forced internal waves reaching the BBL is discussed in section \ref{section:conclusions}, and left for future work.

The total background flow (geostrophic plus wave) in the BBL is then defined to be
\begin{subequations}\label{eq:bckgrnd}
    \begin{equation}
         U(z,t) = \breve{u}_z(t)(z-H),
    \end{equation}
    \begin{equation}
        V(z,t)=V_\infty + \Lambda(z-H) +\breve{v}_z(t)(z-H), \text{ and}
    \end{equation}
    \begin{equation}
        B(x,z,t) = N^2_\infty (\theta x + z) - N^2_\infty \gamma (z-H)+\breve{b}_z(t)(z-H).
    \end{equation}
\end{subequations}
Here time-dependent terms denoted with a breve are associated with the wave and other terms are associated with the steady arrested Ekman solution. Subscript $z$ are used for clarity to identify quantities involving slope-normal gradients.

\begin{figure}[t!]
    \centering
    \includegraphics[width=0.8\linewidth]{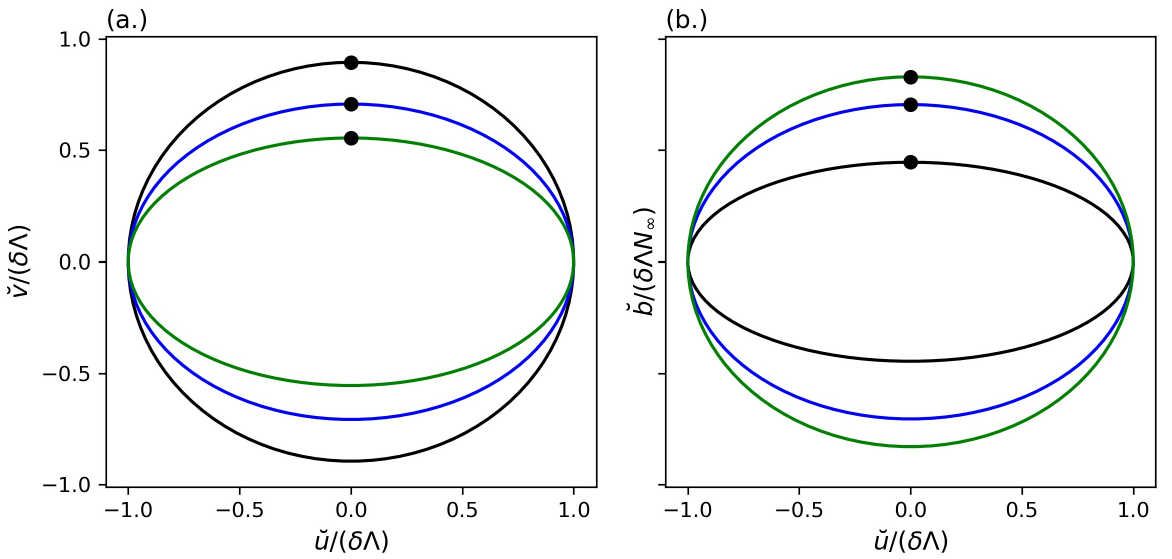}
    \caption{Hodographs for (a.) $\breve{u}_z$-$\breve{v}_z$ and (b.) $\breve{u}_z$-$\breve{b}_z$. The case where $S_\infty=0.5$ is denoted by the black lines, $S_\infty=1.0$ by the blue lines, and $S_\infty=1.5$ by the green lines. In each case the black dot denotes the initial condition.}
    \label{fig:hodographs}
\end{figure}

The inviscid nonhydrostatic Boussinesq equations in the rotated coordinate system are
\begin{subequations}\label{eq:boussinesq}
    \begin{equation}
        \dfrac{Du}{Dt}-fv =  -\dfrac{\partial p}{\partial x}  + \theta b,
    \end{equation}
    \begin{equation}
        \dfrac{Dv}{Dt} + fu - f\theta w = 0,
    \end{equation}
    \begin{equation}
       \frac{Dw}{Dt}+ f\theta v =-\dfrac{\partial p}{\partial z}  + b,
    \end{equation}
    \begin{equation}
        \dfrac{\partial u}{\partial x}+\dfrac{\partial w}{\partial z} = 0, \text{ and}
    \end{equation}
    \begin{equation}
        \dfrac{Db}{Dt} = 0,
    \end{equation}
\end{subequations}
where the flow is assumed uniform in the along-slope direction and $D/Dt:= \partial/\partial t+u\partial/\partial x+w\partial/\partial z$. Using Equation \ref{eq:bckgrnd} in \ref{eq:boussinesq} gives the equations for oscillations in shear and stratification in the arrested Ekman layer \citep{ Brink_Lintz_2010b}
\begin{subequations}\label{eq:shear_oscillations_diff_eq}
    \begin{equation}
        \dfrac{\partial \breve{u}_z}{\partial t}  = f\breve{v}_z+\theta\breve{b}_z,
    \end{equation}
    \begin{equation}
        \dfrac{\partial \breve{v}_z}{\partial t}  = -f\breve{u}_z, \text{ and}
    \end{equation}
    \begin{equation}
        \dfrac{\partial \breve{b}_z}{\partial t}  = -N^2_\infty\theta \breve{u}_z.
    \end{equation}
\end{subequations}
Equation \ref{eq:shear_oscillations_diff_eq} is only defined for the arrested Ekman layer, i.e. $z<H$, as the interior flow is steady and spatially uniform. The initial conditions are given by

\begin{equation}\label{eq:intlcndtns}
    \breve{u}_z(0) = \delta\Lambda\cos\phi, \quad \breve{v}_z(0) = -\dfrac{\delta\Lambda f}{f^*}\sin\phi, \text{ and } \quad \breve{b}_z(0) = -\dfrac{N^2_\infty\theta\delta \Lambda}{f^*}\sin\phi
\end{equation}
where $\phi$ is the initial phase of the wave, and  $\delta$ is the geostrophic shear scaling factor that sets the strength of the near-inertial oscillations relative to the thermal-wind shear. With this, the solutions for the wave shear and stratification are
\begin{subequations}\label{eq:bckgrndsolutions}
    \begin{equation}
        \breve{u}_z(t) = \delta\Lambda\cos(f^*t+\phi),
    \end{equation}
    \begin{equation}
        \breve{v}_z(t) = -\dfrac{\delta\Lambda f}{f^*}\sin(f^*t+\phi), \text{ and}
    \end{equation}
    \begin{equation}
        \breve{b}_z(t) = -\dfrac{N^2_\infty\theta\delta \Lambda}{f^*}\sin(f^*t+\phi).
    \end{equation}
\end{subequations}
These describe oscillations at a modified inertial frequency $f^*= f\sqrt{1+S_\infty^2}$. The dependence on the slope Burger number reflects the importance of buoyancy anomalies generated by across-slope advection relative to Coriolis accelerations in the across-shelf momentum equation, (\ref{eq:shear_oscillations_diff_eq}a). We highlight this as a key difference of the BBL relative to the surface mixed-layer case. While we refer to these oscillations as `near-inertial' (as they are the natural frequency of slope-aligned motions in the BBL) we note that these frequencies will be larger than the local inertial frequency, particularly in the case of large slope Burger numbers (steep topography or strong stratification).

The eccentricities of the oscillations formed by $\breve{u}_z$-$\breve{v}_z$ and $\breve{u}_z$-$\breve{b}_z$ are
\begin{subequations}
    \begin{equation}\label{eq:uvecc}
        e_{\breve{u}_z-\breve{v}_z}=\sqrt{\dfrac{S^2_\infty}{1+S^2_\infty}}, \text{ and }
    \end{equation}
    \begin{equation}\label{eq:ubecc}
        e_{\breve{u}_z-\breve{b}_z} = \sqrt{\dfrac{1}{1 + S^2_\infty}},
    \end{equation}
\end{subequations}
where the velocity shears are normalized by $\delta \Lambda$ and the buoyancy gradient by $\delta\Lambda N_\infty$ to reflect their contributions to the energetics.
If the eccentricity is near $0$, this indicates a circular hodograph in $\breve{u}_z$-$\breve{v}_z$, and if the eccentricity is near $1$, this indicates an elongated, rectilinear hodograph along the x direction (Fig. \ref{fig:hodographs}). The slope-parallel oscillations used here become increasingly rectilinear for increasing $S_\infty$, a consequence of exchanges between kinetic and potential energy due to across-slope motions generating buoyancy anomalies. This can be contrasted with pure inertial oscillations (the limit of $S_\infty\rightarrow 0$), which are circularly polarized in velocity ($e_{\breve{u}_z-\breve{v}_z}=0$) with no exchange between kinetic and potential energy ($e_{\breve{u_z}-\breve{b}_z} = 1$).

For completeness we also include the vertically-averaged total energy equation for the near-inertial oscillations, which is given by
\begin{subequations}\label{eq:background_oscillation_energy}
    \begin{equation}
        \dfrac{\partial E}{\partial t}=0, \text{ and}
    \end{equation}
    \begin{equation}\label{eq:background_oscillation_energy_osc_only}
        E = \frac{1}{H}\int_0^H\left({K}+{P}\right)\;\mathrm{dz}=\dfrac{(\delta V_\infty)^2}{6},
    \end{equation}
\end{subequations}
where ${K}:=(\breve{u}_z^2+\breve{v}_z^2)(z-H)^2/2$ is the wave kinetic energy, and ${P}:=\breve{b}_z^2(z-H)^2/(2N^2_\infty)$ is the wave potential energy. Thus, the vertically-averaged wave energy is a function only of the interior velocity and the relative strength of the wave, $\delta$. The integrated total energy does though have an implicit dependence on the strength of the geostrophic shear via the boundary layer height $H$.


\subsection{Parametric subharmonic instability in the weak-wave limit}\label{section:psi_discussion}
It is informative to consider the parameter dependence of PSI in the weak-wave limit, i.e. $\delta\rightarrow 0$. In this limit, PSI is a weakly nonlinear wave-wave interaction that occurs when the waves' properties constitute a triad interaction. Generally, near-inertial waves are more susceptible to PSI because of their low vertical group velocity \citep{MacKinnonWinters05}, however the instability also requires the local minimum frequency to be sufficiently low as to allow the subharmonic waves to exist.
One way to achieve that is the presence of strong baroclinicity in a flow which modifies the Ertel PV, and hence the minimum allowable wave frequency \citep[in the hydrostatic or strongly stratified limit][]{hilditch_trapping_2026}, as $\omega^2_{min} = f q/N^2$ where $q = (f\hat{k} + \nabla \times \vec{u})\cdot \nabla b$ is the Ertel PV \citep{Whitt_and_Thomas_2013}. Conditions where the PV is lower than the planetary PV, $q<fN^2$, thus have a minimum wave frequency below $f$. We note that once the PV is opposite signed from $f$ (ie. negative in the Northern hemisphere) due to baroclinicity the flow becomes unstable to SI which can be the fastest growing instability \citep{HaineandMarshal1998}, although some cases with strong waves and weakly negative PV can still be dominated by PSI \citep[as in][and below]{Hilditch_Thomas_2023}. 

\begin{figure}[t!]
    \centering
    \includegraphics[width=0.9\textwidth]{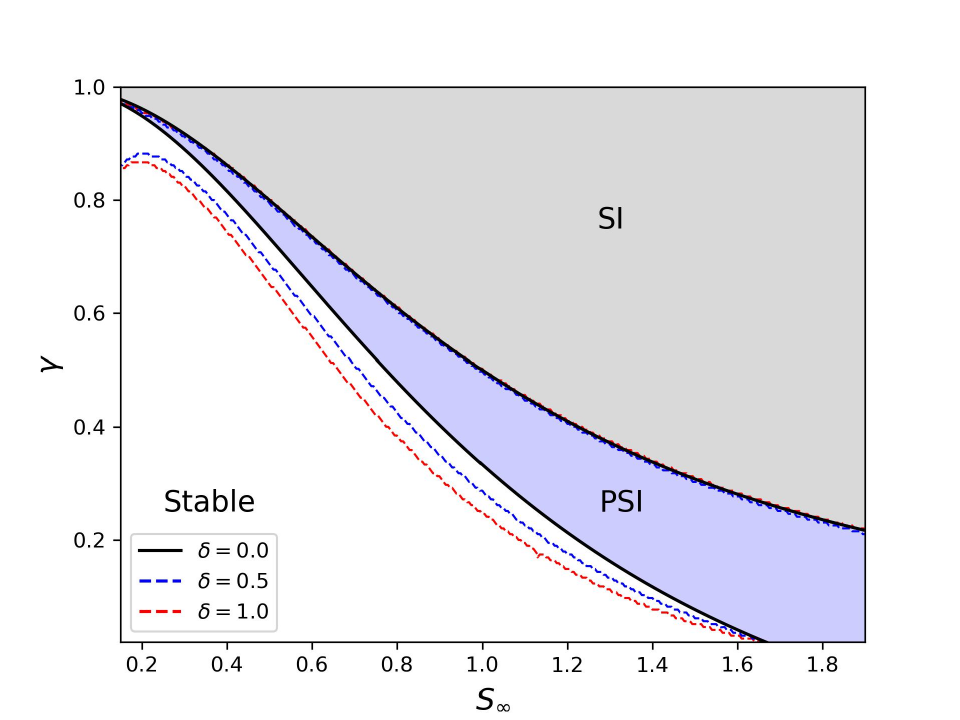}
    \caption{Parameter space in $S_\infty-\gamma$ for PSI, where the blue regions denote the PSI unstable portion of the parameter space, the grey regions are unstable to SI and the white regions are stable to both. The blue and red lines denote the expansion of the PSI unstable region for wave magnitudes of $\delta=0.5$ and $\delta=1.0$ from the Floquet analysis.}
    \label{fig:instability_region}
\end{figure}

For the background flow defined here the total PV is given by
\begin{equation}\label{eq:PV}
    q = f N_\infty^2\left(1-\gamma(1+S^2_\infty)\right),
\end{equation}
which varies relative to the interior PV as a function of the stratification parameter $\gamma$, and the slope Burger number. We note that this is the same as the PV for the arrested Ekman solution without a near-inertial oscillation \citep{Allen_Newberger_1998}, as the initial shear and buoyancy anomalies are phased so as to cancel the wave contribution to the PV. The minimum allowable frequency of inertia-gravity waves in this flow can be written as 
\begin{equation}\label{eq:minfreq}
    \omega^2_{\text{min}}=\dfrac{f q}{N^2}=f^2\left(1 - S^2_\infty\frac{\gamma}{1-\gamma}\right).
\end{equation}
For PSI to take place in the flow while excluding the possibility of SI, Equation \ref{eq:minfreq} must satisfy the inequality
\begin{equation}\label{eq:freqlimits}
    0\leq\omega^2_{\text{min}}\leq \dfrac{(f^*)^2}{4}.
\end{equation}
From this relation, limits on the choice of $\gamma$ can be derived:
\begin{equation}\label{eq:gmlimits}
    \gamma^*_l=\dfrac{3-S^2_\infty}{3(1+S^2_\infty)}\leq \gamma \leq \dfrac{1}{1+S^2_\infty}=\gamma^*_u.
\end{equation}
If $\gamma < \gamma_l^*$ the flow is stable to PSI and if $\gamma > \gamma_u^*$ the flow is symmetrically unstable (fig. \ref{fig:instability_region}). The region of PSI instability broadens for larger slope Burger numbers, but requires smaller stratification anomalies in order to maintain $q>0$. These bounds should be understood as the limit of $\delta\rightarrow0$ such that the wave does not affect the background conditions felt by the growing instability, and Floquet analysis (below, and see figure \ref{fig:instability_region}) shows that the effective  $\gamma^*_l$ limit expands as $\delta$ increases.  We also note that (\ref{eq:gmlimits}) predicts that for $S_\infty^2>3$ it may be possible to allow PSI even in flows with \textit{positive} PV anomalies ($\gamma<0$). This is a consequence of $f^*$ increasing more rapidly as a function of $S_\infty$ than $\omega_{min}$, however in this part of parameter space the slope-parallel waves are no longer truly `near-inertial' hence we focus only on the portion of parameter space with reduced PV ($\gamma>0$). 

A relation similar to (\ref{eq:gmlimits}) can be derived for the ratio of the minimum frequency to the modified inertial frequency. Dividing Equation \ref{eq:freqlimits} by $(f^*)^2$ the inequality can be written in terms of the stratification parameter and slope Burger number, 
\begin{equation}\label{eq:freq_ratio_params}
    0\leq\dfrac{1-\gamma(1+S_\infty^2)}{(1-\gamma)(1+S_\infty^2)}\leq \dfrac{1}{4}.
\end{equation}
This, along with (\ref{eq:dmnmbrs}) and (\ref{eq:background_oscillation_energy}), indicate that the relevant parameter space can be posed in terms of the interior slope Burger number ($S_\infty$) and stratification parameter ($\gamma$)---which together determine the background PV and hence minimum wave frequency---and the strength of the imposed near-inertial wave ($\delta$). Results of the linear stability analysis and numerical simulations are presented in terms of these parameters below.

\subsection{Linear Stability Analysis}\label{section:linearstabilityanalysis}

Here we closely follow the approach of \citet{Hilditch_Thomas_2023} to derive the linear stability problem. The total velocity and buoyancy fields are defined as

\begin{subequations}
    \begin{equation}
        u_T(x,z,t)=u(x,z,t)+U(z,t),
    \end{equation}
    \begin{equation}
        v_T(x,z,t)=v(x,z,t)+V(z,t),
    \end{equation}
    \begin{equation}
        w_T(x,z,t)=w(x,z,t), \text{ and}
    \end{equation}
    \begin{equation}
        b_T(x,z,t)=b(x,z,t)+B(x,z,t),
    \end{equation}
\end{subequations}
where the lower case variables denote the perturbations and the background fields are defined in (\ref{eq:bckgrnd}). Linearizing (\ref{eq:boussinesq}) gives,

\begin{subequations}\label{eq:boussinesqlin}
    \begin{equation}
        \dfrac{D_o u}{Dt}+\breve{u}_zw-f v =  -\dfrac{\partial p}{\partial x}   + \theta b,
    \end{equation}
    \begin{equation}
        \dfrac{D_o v}{Dt}+(\Lambda+\breve{v}_z)w + f u-f\theta w= 0,
    \end{equation}
    \begin{equation}
        \dfrac{D_o w}{Dt}+f\theta v = -\dfrac{\partial p}{\partial z}   + b,
    \end{equation}
    \begin{equation}
        \dfrac{\partial u}{\partial x}+\dfrac{\partial w}{\partial z} = 0, \text{ and}
    \end{equation}
    \begin{equation}
        \dfrac{D_o b}{Dt}+N^2_\infty\theta u + (N^2_\infty[1-\gamma]+\breve{b}_z)w = 0,
    \end{equation}
\end{subequations}
where $D_o/Dt:=\partial/\partial t + U\partial/\partial x$. Defining the wave displacement, 
\begin{equation}
 \chi(t,z) := \int \breve{u}_z(t)(z-H)\mathrm{d}t = \frac{\delta \Lambda}{f^*}\sin(f^*t + \phi)(z-H),
 \end{equation}
we can assume solutions of the form
\begin{equation}\label{eq:AssumedSolForm}
    \varphi =    
    \tilde{\varphi}(t)
    \exp(ik[x-\chi-\mu_o z]), 
\end{equation}
where $\varphi$ is any variable from Equation \ref{eq:boussinesqlin}. A consequence of this choice of solutions is the result
\begin{equation}
    \dfrac{D_o \varphi}{Dt}=\dfrac{d \tilde{\varphi}}{d t}\exp(ik[x-\chi-\mu_o z]).
\end{equation}
Plugging in these solutions, the equations are transformed to 
\begin{subequations}
    \begin{equation}
        \dfrac{d \tilde{u}}{d t}+\breve{u}_z\tilde{w}-f\tilde{v} =  -ik\tilde{p}  + \theta\tilde{b},
    \end{equation}
    \begin{equation}
        \dfrac{d \tilde{v}}{d t}+(\Lambda+\breve{v}_z)\tilde{w} + f\tilde{u}-f\theta \tilde{w}= 0,
    \end{equation}
    \begin{equation}
        \dfrac{d \tilde{w}}{d t}+f\theta \tilde{v} = ik\left(\dfrac{\partial \chi}{\partial z}+\mu_o\right)\tilde{p}  + \tilde{b},
    \end{equation}
    \begin{equation}\label{eq:continuity}
        \tilde{u} -  \left(\dfrac{\partial \chi}{\partial z}+\mu_o\right)\tilde{w} = 0, \text{ and}
    \end{equation}
    \begin{equation}
        \dfrac{d \tilde{b}}{d t}+N^2_\infty\theta\tilde{u} + (N^2_\infty[1-\gamma]+\breve{b}_z)\tilde{w} = 0.
    \end{equation}
\end{subequations}
Let us define the function
\begin{equation}
    \mu(t):= \mu_o + \frac{\partial \chi}{\partial z} = \mu_o+\dfrac{\delta\Lambda}{f^*}\sin(f^*t+\phi),
\end{equation}
and apply the continuity equation (Equation \ref{eq:continuity}) to get
\begin{subequations}
    \begin{equation}\label{eq:umomenttrans}
        \mu\dfrac{d \tilde{w}}{d t}+2\dfrac{d \mu}{d t}\tilde{w}-f \tilde{v} =  -ik\tilde{p}  + \theta\tilde{b},
    \end{equation}
    \begin{equation}
        \dfrac{d \tilde{v}}{d t} + (\Lambda + f\mu_o-f\theta )\tilde{w}= 0,
    \end{equation}
    \begin{equation}\label{eq:wmomenttrans}
        \dfrac{d \tilde{w}}{d t}+f\theta v = ik\mu p  + \tilde{b}, \text{ and}
    \end{equation}
    \begin{equation}
        \dfrac{d \tilde{b}}{d t} + (N^2_\infty[1-\gamma]+N^2_\infty\theta\mu_o)\tilde{w} = 0.
    \end{equation}
\end{subequations}
Next multiplying Equation \ref{eq:umomenttrans} by $\mu(t)$ and adding it to Equation \ref{eq:wmomenttrans}
\begin{subequations}
    \begin{equation}
        \dfrac{d}{d t}\left([1+\mu^2]\tilde{w}\right)=-(f\theta - f\mu) \tilde{v} +(1+\theta\mu) \tilde{b},
    \end{equation}
    \begin{equation}
        \dfrac{d \tilde{v}}{d t} + (\Lambda + f \mu_o-f\theta )\tilde{w}= 0,
    \end{equation}
    \begin{equation}
        \dfrac{d \tilde{b}}{d t} + (N^2_\infty[1-\gamma]+N^2_\infty\theta\mu_o)\tilde{w} = 0.
    \end{equation}
\end{subequations}
Defining $\zeta:=\int\tilde{w}dt$, $\Psi = (1+\mu(t)^2)\tilde{w}$, and the non-dimensionalization $\hat{t} :=f^* t$ we recover the system of ordinary differential equations
\begin{subequations}\label{eq:stblyeqn}
    \begin{equation}
        \dfrac{d\Psi}{d\hat{t}}= (A_0-A_1\mu(\hat{t}))\zeta, \text{ and }
    \end{equation}
    \begin{equation}
        \dfrac{d\zeta}{d \hat{t}}= (1+\mu(\hat{t})^2)^{-1}\Psi,
    \end{equation}
\end{subequations}


where
\begin{subequations}
    \begin{equation}
        A_0 = \frac{\theta^2 - S_\infty^2}{\theta(1 + S_\infty^2)}\left[\mu_0 - \frac{\theta}{\theta^2 - S_\infty^2}\left(\frac{S_\infty^2}{\theta^2}(1 - \gamma) + \theta^2\right)\right], \text{ and }
    \end{equation}
    \begin{equation}
        A_1 = \mu_0 + \frac{S_\infty^2 - \theta^2}{\theta(1 + S_\infty^2)}.
    \end{equation}
\end{subequations}
Utilizing the small angle approximation that $\theta^2\ll 1$ (except where multiplied by $N^2_\infty/f^2$ as in the slope Burger number), once again these are reduced to 
\begin{subequations}
    \begin{equation}
        A_0 \approx -\frac{S_\infty^2}{\theta(1 + S_\infty^2)}\left[\mu_0 + \frac{1-\gamma}{\theta}\right], \text{ and }
    \end{equation}
    \begin{equation}
        A_1 = \mu_0 + \frac{S_\infty^2}{\theta(1 + S_\infty^2)}.
    \end{equation}
\end{subequations}

Note that this system of equations closely resembles the equations for the stability analysis from \cite{Hilditch_Thomas_2023} considering PSI at a surface front, and indeed under certain conditions the problems can be shown to be mathematically equivalent. 
The inverse Richardson number as defined in \cite{Hilditch_Thomas_2023} has the equivalence
\begin{equation}\label{eq:hilditchthomasrichardsonnumber}
    \text{Ri}_{\text{HT}}^{-1}\equiv \dfrac{S^2_\infty}{(1+S^2_\infty)(1-\gamma)},
\end{equation}
which for this flow is the ratio of the boundary layer stratification that gives $q=0$, to the actual BBL stratification. An important difference here is that while the PV anomaly is fully determined by the Richardson number in a surface frontal zone where there is only vertical shear, here the slope normal shear projects on both the horizontal and vertical components of vorticity (see eg. \ref{eq:dmnmbrs}).

\begin{figure}
    \centering
    \includegraphics[width=0.95\textwidth]{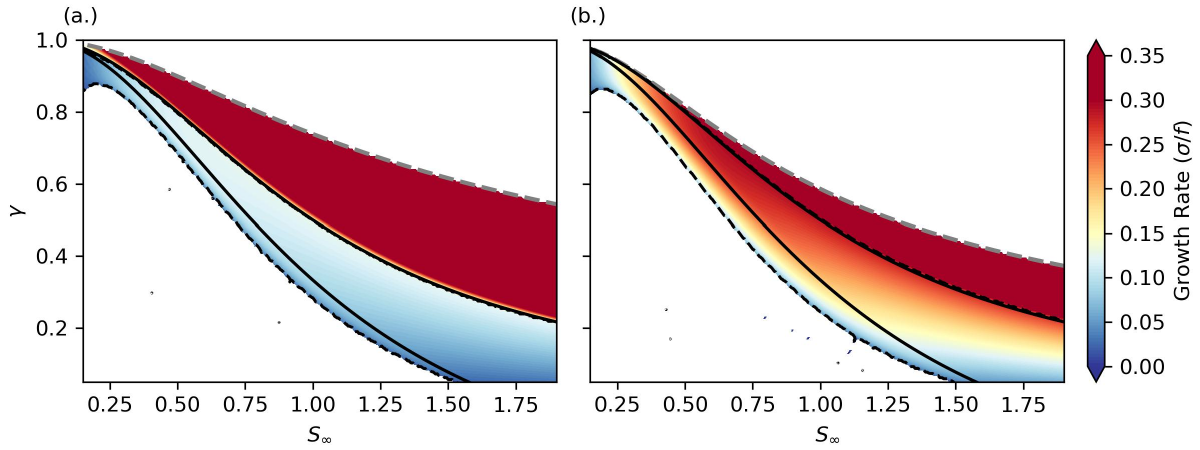}
    \caption{Growth rates over $S_\infty$-$\gamma$ parameter space for (a.) $\delta=0.5$; and (b.) $\delta=1.0$. The solid black lines demarcate regions that are PSI unstable from theory. Portions of the parameter space that are convectively unstable at any phase of the wave (above the gray dashed line), and regions where the imaginary (oscillatory) component of the growth rate does not equal $f^*/2$, indicating it is not PSI, are excluded.}
    \label{fig:grwth_rt_prmtr_space_Sinf_gamma}
\end{figure}

\subsection{Solutions}\label{section:solutions}
The coefficients of Equation \ref{eq:stblyeqn} are periodic and time reversible. Employing Floquet theory, any solution of Equation \ref{eq:stblyeqn} satisfies
\begin{equation}
    \begin{bmatrix}
        \Psi(\hat{t}\,) \\
        \zeta(\hat{t}\,)
    \end{bmatrix} = P_1 e^{\sigma \hat{t}}\begin{bmatrix}
        \Psi_1(\hat{t}\,) \\
        \zeta_1(\hat{t}\,)
    \end{bmatrix} + P_2 e^{-\sigma \hat{t}}\begin{bmatrix}
        \Psi_2(\hat{t}\,) \\
        \zeta_2(\hat{t}\,)
    \end{bmatrix}
\end{equation}
for some choice of coefficients $P_1,P_2$ where $\Psi_1$, $\Psi_2$, $\zeta_1$, and $\zeta_2$ are periodic eigenfunctions \citep{Floquet1883}. To solve for the growth rate, $\sigma$, two linearly independent initial conditions for Equation \ref{eq:stblyeqn} are numerically integrated over a single period and the eigenvalues are solved for. For solutions which produce PSI, the imaginary (oscillatory) component of the complex growth rate will have $\text{Im}(\sigma) = \pm 0.5$ corresponding to the subharmonic frequency.

PSI in the arrested Ekman layer can begin to be explored by inspecting the real components of the Floquet exponents over the $S_\infty-\gamma$ parameter space seen in Figure \ref{fig:grwth_rt_prmtr_space_Sinf_gamma}. The parameter space shows regions where the flow is PSI unstable (middle region), SI unstable (upper region), and stable (lower region). Growth rates tend to be strongest nearest $\gamma_u$, where the background PV approaches $0$ near the SI unstable region. The PSI unstable region expands into the stable portion of the space below $\gamma_l$ as $\delta$ increases, with the largest expansion being for smaller slope Burger numbers where the near-inertial oscillations approach the inertial frequency ($f^*\rightarrow f$). The PSI unstable region expanding into what is predicted to be stable flow regimes from the constraints on the frequency ratio is consistent with what was found for PSI of a surface front in \cite{Hilditch_Thomas_2023}. The region unstable to PSI does not expand as much for increasing slope Burger numbers, with more of the parameter space becoming stable. This may be a result of the increase of $f^*$ for larger $S_\infty$. The growth rates further decrease with increasing slope Burger number near $\gamma_l$, which could be the result of $\Lambda$, the parameter measuring the shear in the background flow, decreasing with $\gamma$.

\begin{figure}
    \centering
    \includegraphics[width=0.95\linewidth]{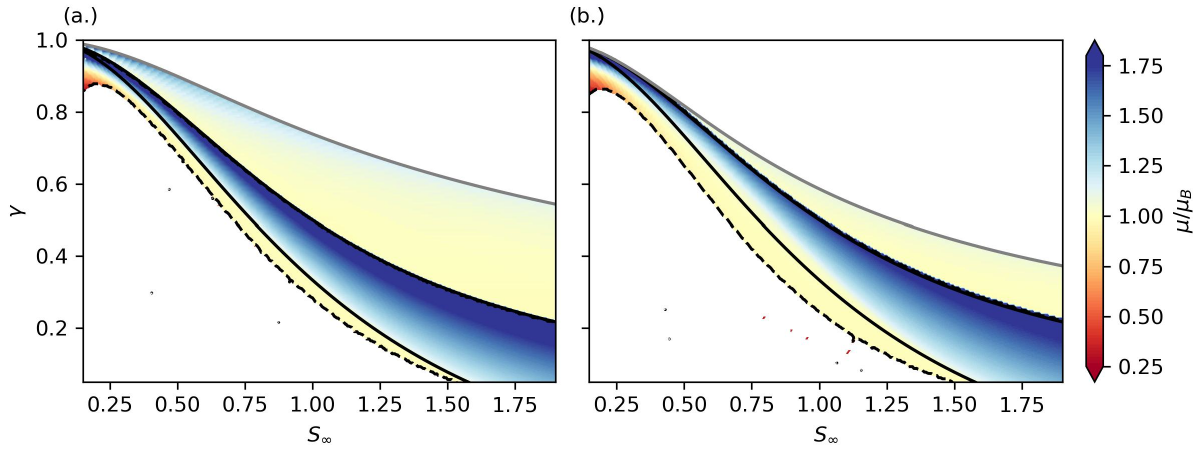}
    \caption{Perturbation slope angles in $S_\infty$-$\gamma$ parameter space for (a.) $\delta=0.5$; and (b.) $\delta=1.0$. The solid black lines demarcate regions that are PSI unstable from theory. Portions of the parameter space that are convectively unstable at any phase of the wave  (above the gray line), and regions where the imaginary (oscillatory) component of the growth rate does not equal $f^*/2$, indicating it is not PSI, are excluded.}
    \label{fig:slp_angl_prmtr_space}
\end{figure}

The perturbation slope angles of PSI as a function of slope Burger number and stratification parameter are shown in Figure \ref{fig:slp_angl_prmtr_space}, normalized such that a value of 1 indicates alignment with the isopcynals of the balanced background state. The perturbations become less steeply inclined, relative to isopycnals, as the SI stability boundary is approached ($\mu/\mu_B>1)$, consistent with previous work on surface fronts \citep{Hilditch_Thomas_2023}. Moving towards $\gamma_l$ the perturbation slope angle becomes more aligned with the mean isopycnal slope, consistent with the subharmonics having a frequency closer to the minimum allowable frequency, and thus perturbations aligned with isopycnals \citep{Whitt_and_Thomas_2013}. Increasing slope Burger number leads to more uniformly shallow perturbation slopes. For a larger geostrophic scaling factor (Figure \ref{fig:slp_angl_prmtr_space}b), the perturbations more closely align with isopycnals, a result also consistent with \cite{Hilditch_Thomas_2023}. 

\begin{figure}
    \centering
    \includegraphics[width=0.95\textwidth]{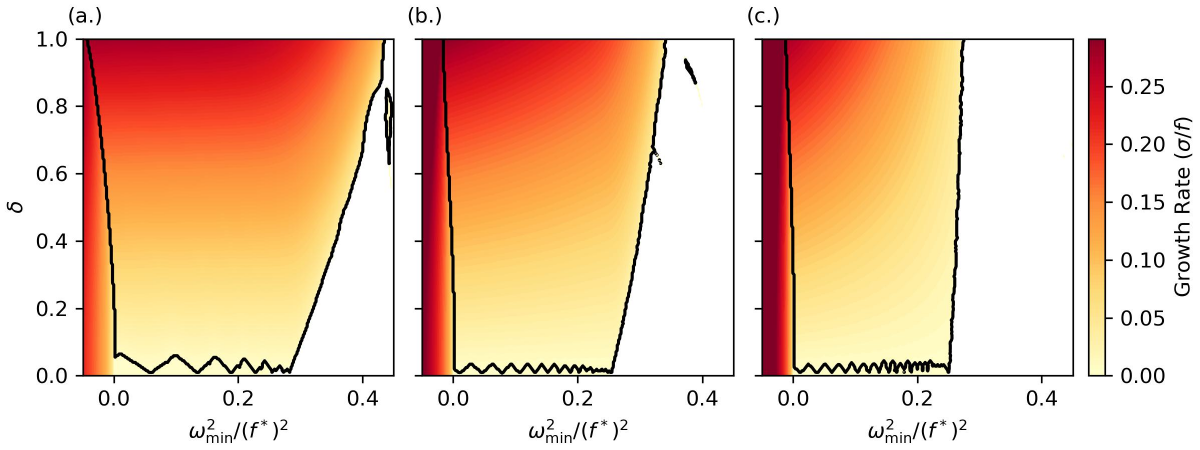}
    \caption{Growth rates over $\omega_{\text{min}}^2/(f^*)^2-\delta$ parameter space for (a.) $S_\infty=0.5$; (b.) $S_\infty=1$; and (c.) $S_\infty=1.5$. The black lines demarcate regions that are PSI unstable. Note that portions of the parameter space where $\omega_{\text{min}}^2/(f^*)^2>0.25$, and regions where the imaginary (oscillatory) component of the growth rate does not equal $f^*/2$ are filtered out to remove noise.}
    \label{fig:grwth_rt_prmtr_space}
\end{figure}
For completeness, and to facilitate comparison with prior work, we also can define the parameter space in terms of the frequency ratio, $(\omega_{min}/f^*)^2$, and the wave magnitude, $\delta$ (Fig. \ref{fig:grwth_rt_prmtr_space}). The growth rates do not vary strongly across frequency ratio values for small wave magnitudes. The wave magnitude in general exerts a strong control on the growth rate for the PSI unstable region, as expected from the $\delta^2$ dependence of the average wave energy (\ref{eq:background_oscillation_energy_osc_only}). The PSI unstable region is the largest for smaller slope Burger numbers, which is consistent with the results presented in $S_\infty-\gamma$ parameter space. These results are broadly consistent with \citet[see their Fig. 2]{Hilditch_Thomas_2023}.

To examine the perturbation kinetic energy (TKE) budget we non-dimensionalize the perturbation velocities as $\tilde{u}:=\Lambda H \hat{u}$, $ \tilde{v}:=\Lambda H \hat{v}$, and $ \tilde{w}:= f^*H \hat{w}$. Further, the non-dimensionalized buoyancy is defined by $\tilde{b}:=N^2_\infty(1-\gamma)H\hat{b}$ and time is non-dimensionalized as defined previously. Applying these non-dimensionalizations, the TKE budget for the linear solutions is given by
\begin{equation}\label{eq:TKE_stability_analysis}
    \dfrac{\partial \hat{k}}{\partial \hat{t}}= -\underbrace{\hat{v}\hat{w}}_{\text{GSP}} -  \underbrace{\delta\cos(\hat{t}+\phi)\hat{u}\hat{w}+\delta \dfrac{f}{f^*}\sin(\hat{t}+\phi)\hat{v}\hat{w}}_{\text{ASP}}+\underbrace{\dfrac{1-\gamma}{\gamma \sqrt{1+S^2_\infty}}\hat{u}\hat{b} + \text{Ri} \hat{w}\hat{b}}_{\text{BP}}
\end{equation}
where $\hat{k}:=(\hat{u}^2+\hat{v}^2+(f^*/\Lambda)^2\hat{w}^2)/2$. The shear production terms are decomposed into geostrophic shear production (GSP) and ageostrophic shear production (ASP) due to near-inertial oscillations \citep{ThomasTaylor14}. Vertical buoyancy production (BP) releases potential energy, and consists of both across-slope and slope-normal components. The across-slope BP is modulated by slope Burger number and stratification parameter, and the slope-normal direction is modulated by the Richardson number.

\begin{figure}
    \centering
    \includegraphics[width=0.95\linewidth]{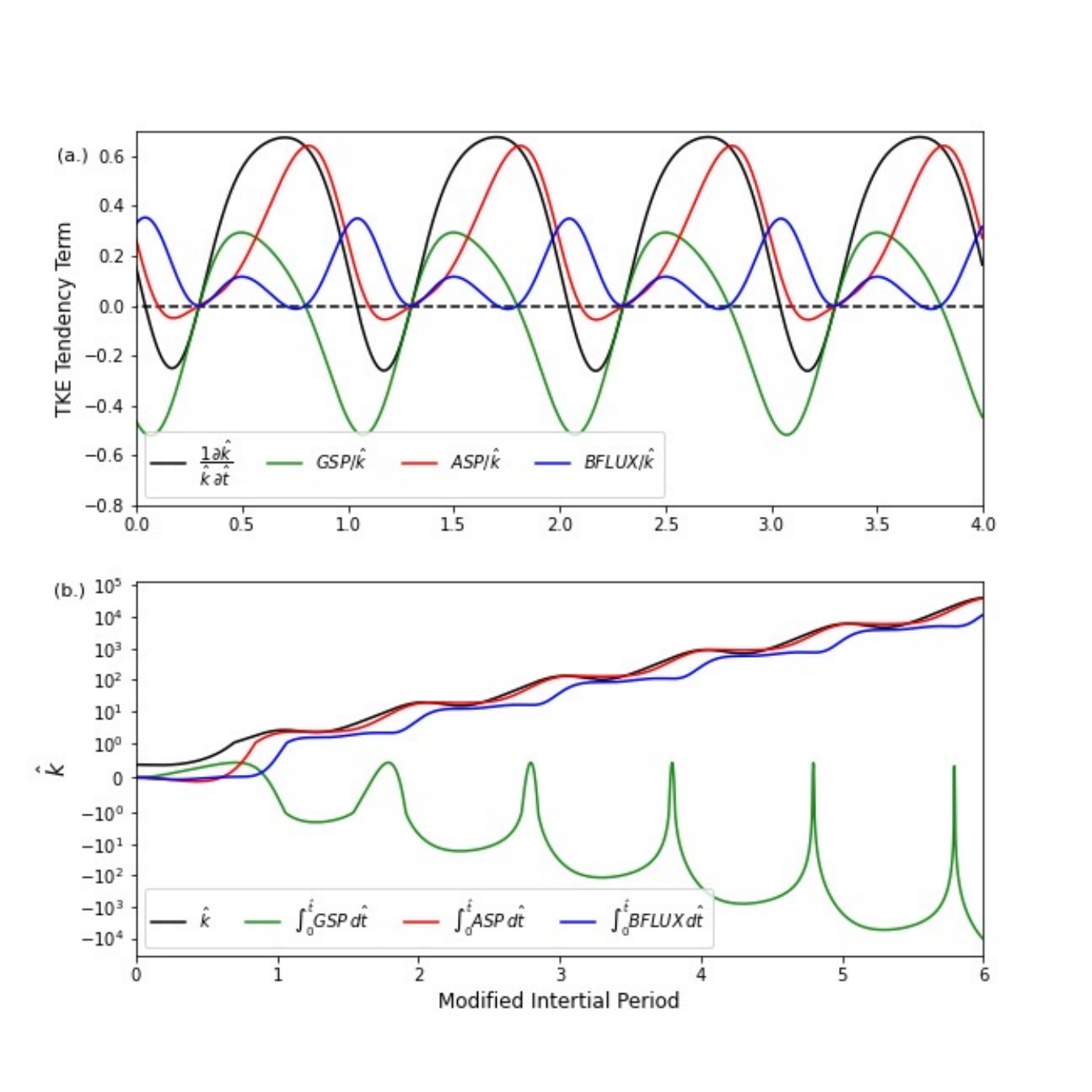}
    \caption{The TKE budget calculated from Equation \ref{eq:stblyeqn} for $S_\infty=1$, $\delta=0.8$, and $\gamma=\gamma_u$ where the modified inertial period is $f^*t/(2\pi)$. (a.) The TKE tendency terms over $4$ inertial periods. (b.) The time integrated TKE tendency terms.}
    \label{fig:stblty_TKE_example}
\end{figure}

The general properties of the TKE budget are shown in Figure \ref{fig:stblty_TKE_example}. TKE growth is driven by ASP, which oscillates at $f^*$ but is almost always positive (Fig. \ref{fig:stblty_TKE_example}a). BP and GSP largely compensate one another in the time-integrated budget, although the GSP varies primarly at the modified inertial frequency whereas BP has variability at both the first and second harmonic. GSP exerts a strong control on the TKE tendency in the case shown, with largest magnitude values when the oscillation is aligned with the mean flow (such that  $\breve{u}_z$ is near 0, fig. \ref{fig:hodographs}). In this inviscid analysis GSP provides the largest integrated sink of energy, however numerical results (section \ref{section:Simulations}) indicate that dissipation also plays an important role in the instability energetics.

\begin{figure}
    \centering
    \includegraphics[width=0.95\linewidth]{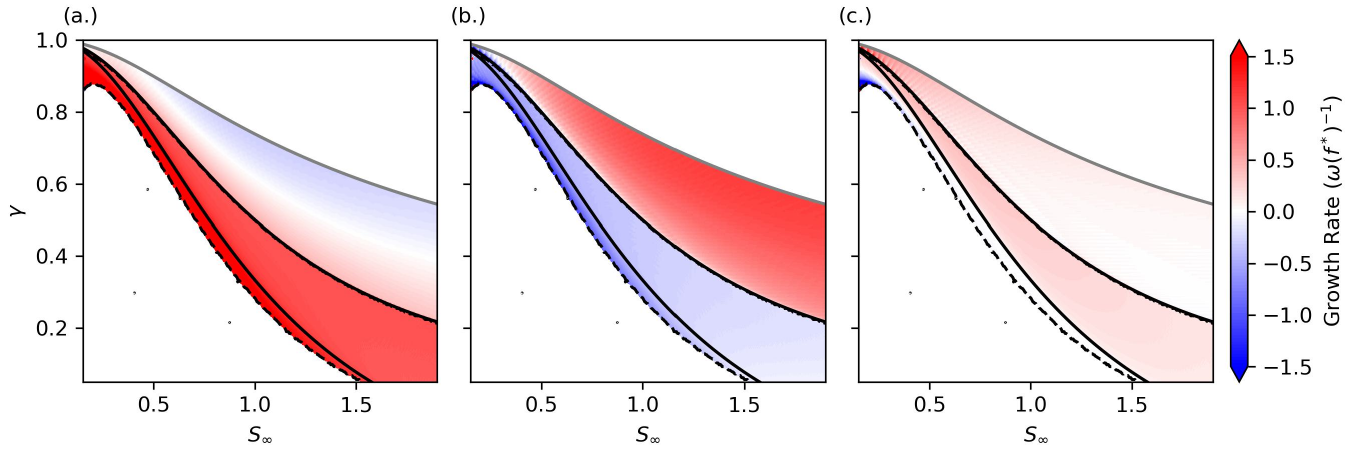}
    \caption{Relative contributions of the (a.) ASP, (b.) GSP, and (c.) BP to the growth of TKE for $\delta=0.5$. The contributions are calculated by integrating TKE term over a single inertial period and dividing by the change in TKE.}
    \label{fig:TKE_rel_contr}
\end{figure}

The parameter dependence of the energy production terms is shown in figure \ref{fig:TKE_rel_contr}.
The ASP is largest for $\gamma$ values near $\gamma_l$ (approaching stability).
BP has the strongest positive contributions for $\gamma$ near $\gamma_u$, whereas GSP removes the most energy for $\gamma$ near $\gamma_l$ when perturbations are aligned closely with isocpynals (Fig. \ref{fig:TKE_rel_contr}b). The sum of $GSP + BP$ is shown for several values of $S_\infty$ in figure \ref{fig:BP_GSP_rel_contr}. The BP contributions cancel the GSP sink of energy for the region between $\gamma_l$ and $\gamma_u$, with increasing cancellation for larger slope Burger numbers. In sum the $GSP + BP$ however lead to a removal of energy for PSI, although we emphasize that this is always smaller than the positive source of TKE from $ASP$.


\section{Simulations}\label{section:Simulations}
\subsection{Numerical Model}\label{section:numericalmodel}

The flow developed was simulated in the Julia package Oceananigans.jl \citep{wagner2025highlevelhighresolutionoceanmodeling} solving the non-hydrostatic, incompressible Navier-Stokes equations under the Boussinesq approximation. Oceananigans.jl solves for the velocity and buoyancy fields for the initial conditions provided and uses a finite volume method on a staggered grid with a 2nd order centered difference advective scheme. Time steps are performed using a 3rd order Runge-Kutta method with a step that varies based on the CFL number. Oceananigans.jl is run using a resolution of $\Delta x\approx 1.95 \text{ m}$ by $\Delta z \approx 0.78\text{ m}$ for a domain that is periodic in the $x$-direction. Simulations are run on a horizontal domain of size $L_x=2000 \text{ m}$, and a vertical domain length was chosen to keep the top of the boundary layer sufficiently separated from the top of the domain (Table \ref{tab:simulation_parameters}). The velocity field has free-slip boundary conditions at $z=0, L_z$ where $L_z$ denotes the top of the domain. The boundary conditions on the buoyancy gradient are chosen to match the background conditions, $\partial b/ \partial z=0$ at $z=0, L_z$. The equations are solved in the rotated frame by rotating the gravity and planetary rotation vectors by the angle $\theta=0.1$ (noting we do not make the small-angle approximation here, cf. section \ref{section:theoryanalysis}), and imposing a uniform interior stratification \citep[as in][]{Wenegrat_et_al_2020}. The Coriolis parameter is assumed to be a constant value of $f=10^{-4} \,\text{s}^{-1}$ to be consistent with the values near the mid-latitudes. Viscosity and diffusivity are assumed constant with the value of $\nu=\kappa=10^{-5}\, \text{m}^{2}\, \text{s}^{-1}$.  The arrested Ekman layer with the added near-inertial oscillations derived earlier are imposed as background fields on the model. All simulations shown here are run for the initial conditions of phase $\phi=-\pi/2$, however other phases were tested and found to not significantly affect results. The velocity fields are initialized with weak random noise (standard deviation of $10^{-6}V_\infty$) to ensure a sufficiently long linear growth period before finite amplitude is reached. To implement the transition between the background conditions in the interior and the boundary layer, the Heaviside function is utilized in the approximate form $\hat{H}(s) \approx 0.5(1+\tanh(s\,10^{4}\;\mathrm{m^{-1}}))$ as it was found to be more efficient for simulations on GPUs.

\begin{figure}
    \centering
    \includegraphics[width=0.95\linewidth]{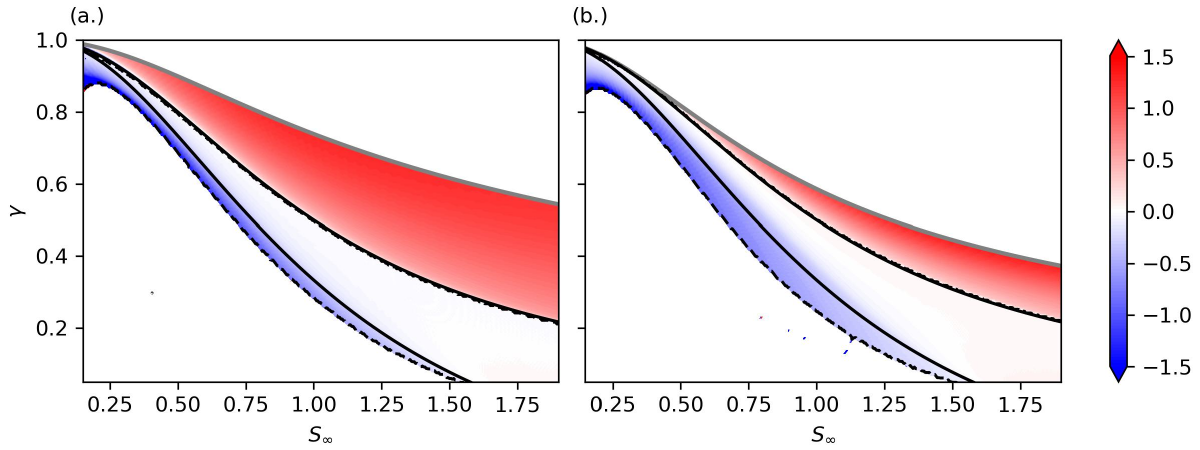}
    \caption{The relative contributions of GSP+BP to the growth of TKE for (a.) $\delta=0.5$; and (b.) $\delta=1.0$. The contributions are calculated by integrating TKE term over a single inertial period and dividing by the change in TKE. }
    \label{fig:BP_GSP_rel_contr}
\end{figure}

The perturbation quantities are defined as
\begin{subequations}
    \begin{equation}
        \vec{u}'=\vec{u}-\overline{\vec{u}},
    \end{equation}
    \begin{equation}
        b'=b-\overline{b}, \text{ and}
    \end{equation}
    \begin{equation}
        p'=p-\overline{p},
    \end{equation}
\end{subequations}
where the operator $\overline{\,\cdot\,}$ is the average in the $x$-plane. The TKE budget is calculated using the following TKE tendency equation:
\begin{equation}
    \begin{split}
        \dfrac{\partial k}{\partial t}= -\underbrace{\overline{v'w'}\Lambda}_{\text{GSP}}-\underbrace{\overline{u'w'}\dfrac{\partial \overline{u}}{\partial z}-\overline{v'w'}\dfrac{\partial \overline{v}}{\partial z} - \overline{u'w'}\breve{u}_z(t)-\overline{v'w'}\breve{v}_z(t)}_{\text{ASP}}+\underbrace{\overline{u'b'\sin\theta+ w'b'\cos\theta}}_{\text{BP}} \\
        -\underbrace{\dfrac{\partial}{\partial z}\left(\overline{w'k'}+\overline{w'p'}-\nu\dfrac{\partial k}{\partial z}\right)}_{\text{TRANS}}-\underbrace{\epsilon}_{\text{DISS}},
    \end{split}
\end{equation}
where $k:=\overline{u'^2+v'^2+w'^2}/2$ is TKE. Note that GSP and BP are the same terms as previously defined in Equation \ref{eq:TKE_stability_analysis}, however here ASP also includes a contribution from other possible ageostrophic shears beyond the imposed wave. TKE transport (TRANS) is $0$ upon volume integrating due to the periodic domain and free-slip boundary conditions. TKE dissipation (DISS) corresponds to the removal of TKE from the flow over time and is defined as $\varepsilon:= \sum_{i=1}^3 \sum_{j=1}^2\nu \overline{\partial u_i/\partial x_j \partial u_i/\partial x_j}$ where $(u_1,u_2,u_3)=(u,v,w)$ and $(x_1,x_2)=(x,z)$. 

\subsection{Simulation Output} \label{section:simulationoutput}

\begin{table}
  \begin{center}
\def~{\hphantom{0}}
  \begin{tabular}{lcccccc}
      Simulation & $S_\infty$ & $\omega^2_{\text{min}}/(f^*)^2$ & $\gamma$ & $H [\text{m}]$ & $L_z [\text{m}]$ & $\delta$\\
      \hline
       Sim1 & 0.5 & 0.35 & 0.692 & 290.8 & 400 & 0.8 \\
       Sim2 & 0.5 & 0.25 & 0.733 & 274.6 & 400 & 0.8 \\
       Sim3 & 0.5 & 0.15 & 0.765 & 263.3 & 400 & 0.8 \\
       Sim4 & 0.5 & 0.05 & 0.789 & 255.0 & 400 & 0.8 \\
       Sim5$^*$ & 0.5 & -0.05 & 0.810 & 248.7 & 400 & 0.8 \\
       Sim6 & 1.0 & 0.35 & 0.231 & 218.1 & 300 & 0.8 \\
       Sim7 & 1.0 & 0.25 & 0.333 & 151.0 & 250 & 0.8 \\
       Sim8 & 1.0 & 0.15 & 0.412 & 122.2 & 200 & 0.8 \\
       Sim9 & 1.0 & 0.05 & 0.474 & 106.3 & 200 & 0.8 \\
       Sim10$^*$ & 1.0 & -0.05 & 0.524 & 96.1 & 200 & 0.8 \\
       Sim11 & 1.5 & 0.25 & 0.0769 & 290.8 & 400 & 0.8 \\
       Sim12 & 1.5 & 0.15 & 0.186 & 120.6 & 200 & 0.8 \\
       Sim13 & 1.5 & 0.05 & 0.271 & 82.5 & 200 & 0.8 \\
       Sim14$^*$ & 1.5 & -0.05 & 0.341 & 65.7 & 150 & 0.8 \\
       Sim15 & 1.0 & 0.2 & 0.375 & 134.2 & 200 & 0.8 \\
       Sim16 & 1.0 & 0.2 & 0.375 & 134.2 & 200 & 0.6 \\
       Sim17 & 1.0 & 0.2 & 0.375 & 134.2 & 200 & 0.4 \\
       Sim18 & 1.0 & 0.2 & 0.375 & 134.2 & 200 & 0.2 \\
  \end{tabular}
  \caption{Slope Burger number, frequency ratio, stratification parameter, boundary layer height and vertical domain size. Note that the asterisk denotes SI unstable simulations.} 
  \label{tab:simulation_parameters}
  \end{center}
\end{table}

The simulation parameters are provided in Table \ref{tab:simulation_parameters} where the BBL height, frequency ratio, geostrophic scaling factor, and slope Burger number are listed. These were chosen to represent a wide sampling of the parameter space for the bottom boundary layer and PSI \citep{ThomasTaylor14,Hilditch_Thomas_2023}. The interior velocity is fixed to $V_\infty = 0.05 \text{ m s}^{-1}$ for most simulations. The geostrophic shear scaling parameter was fixed to a value of $\delta = 0.8$ for most simulations, but select experiments were run that deviated from this value. 

\begin{figure}
    \centering
    \includegraphics[width=0.75\linewidth]{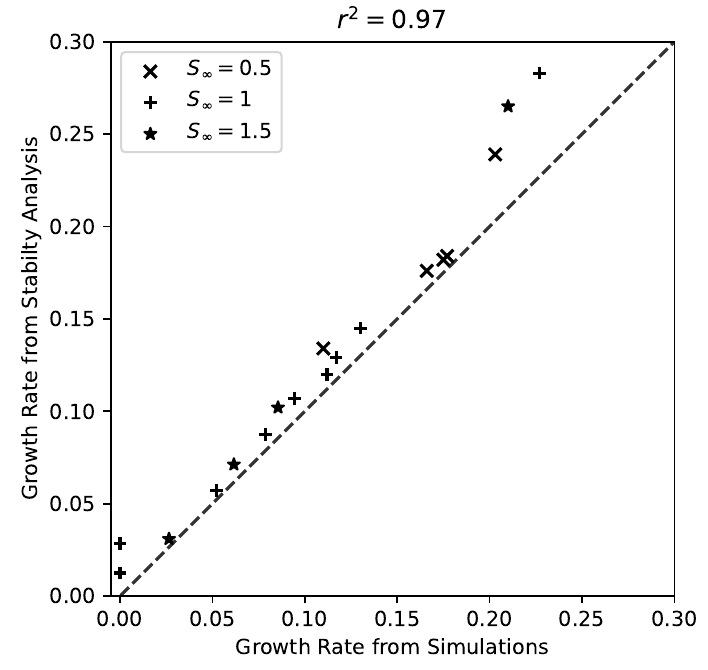}
    \caption{Comparison between the growth rate from the linear stability analysis and the growth rate from the linear phase of the full simulation for the PSI unstable portion of the parameter space. The squared Pearson correlation coefficient is plotted at top.}
    \label{fig:sim_vs_stblty}
\end{figure}

The growth rate estimated from the linear phase of the simulations and the growth rate from linear stability analyses follow each other closely (Fig. \ref{fig:sim_vs_stblty}). Predicted growth rates from the linear stability analysis slightly exceed the numerical solutions, particularly for simulations with faster growth rates, likely due to the inviscid assumption used in Section \ref{section:linearstabilityanalysis}. The flow properties and energetics of Sim9 will be explored first as a representative example, followed by brief discussion of Sim7 for comparison.

\begin{figure}[t!]
    \centering
    \includegraphics[width=0.95\linewidth]{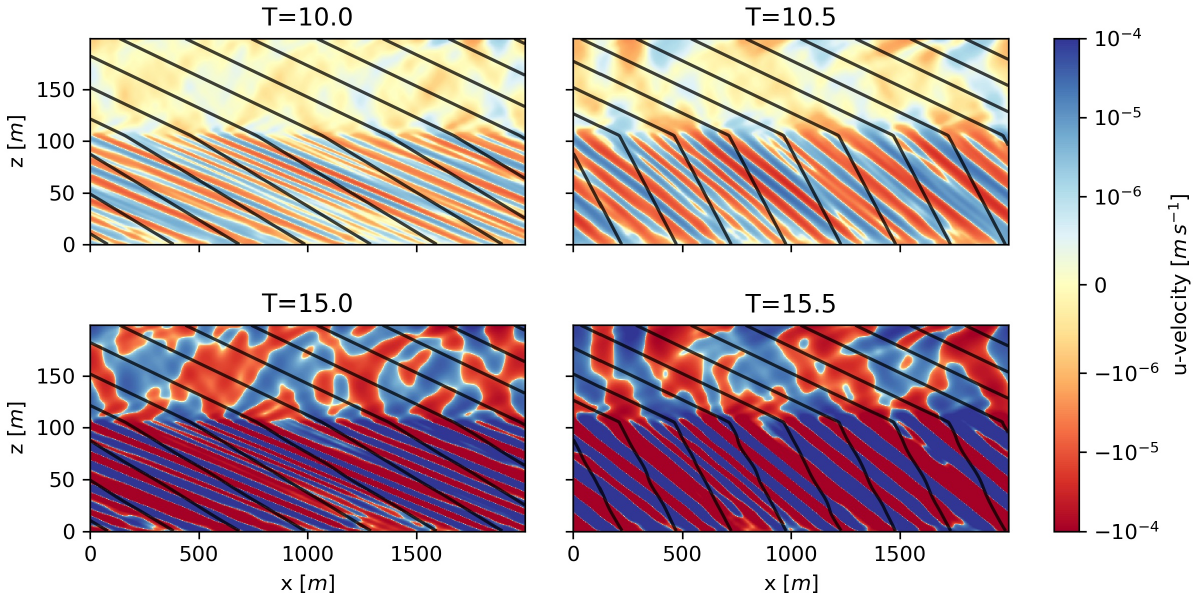}
    \caption{The $u$ perturbation velocities over the linear growth phase of PSI for Sim9, with isopcynals shown in black. The color bar uses a symmetric logarithmic scale, with a linear threshold at $10^{-6}$ m s$^{1}$. The time is normalized by the modified inertial period.}
    \label{fig:S1Rinv095delta08linearphase}
\end{figure}

\begin{figure}
    \centering
    \includegraphics[width=0.95\linewidth]{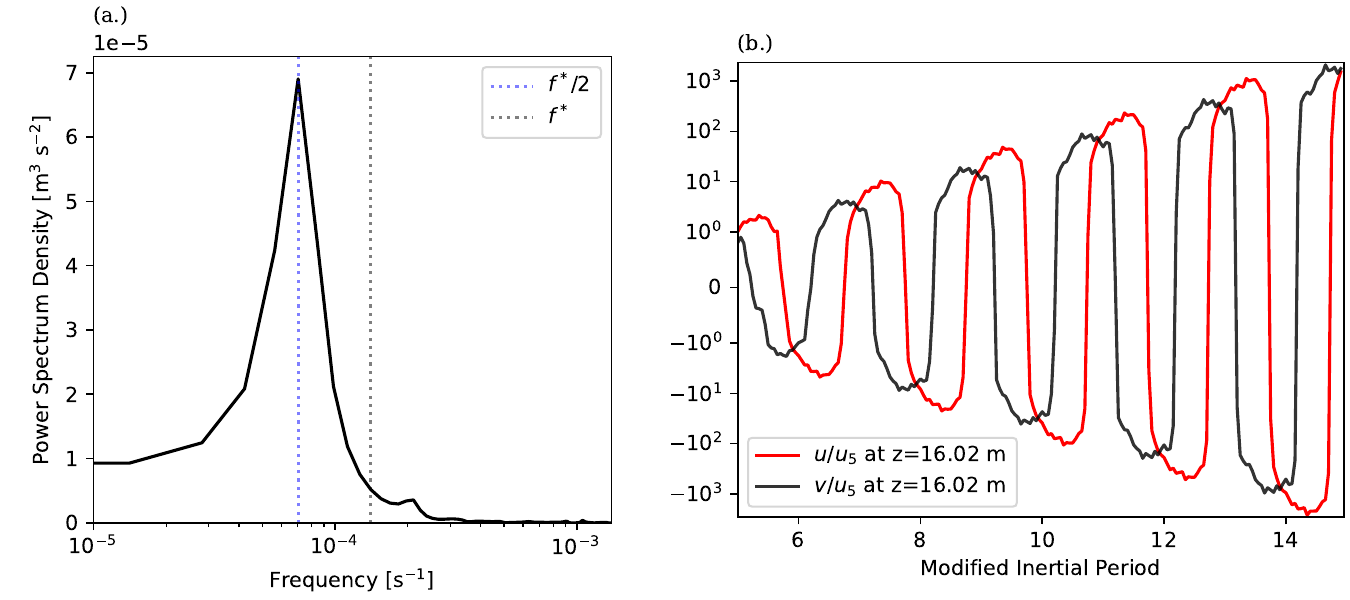}
    \caption{Sim9: (a.) The power spectral density of the $u$ velocity in the the quasi-Lagrangian coordinate-frame, (\ref{eq:QuasiLagrangianCoord}). (b.) The perturbation velocities, again in the quasi-Lagrangian frame, normalized to the value after 5 modified inertial periods. The normalized velocity is plotted on a symmetric logarithm scale, with a linear threshold at 1 and the modified inertial period is $f^*t/(2\pi)$.}
    \label{fig:S1Riinv095delta08fouriertransform}
\end{figure}

The linear growth phase of PSI can be seen in (Fig. \ref{fig:S1Rinv095delta08linearphase}). The instability is advected by the oscillating background flow, but always has a shallower angle than the isopycnals, consistent with expectations from linear stability analysis (Fig. \ref{fig:slp_angl_prmtr_space}). To calculate the frequency spectrum of the subharmonic wave we define a quasi-Lagrangian coordinate following the background isopycnals as 
\begin{equation}\label{eq:QuasiLagrangianCoord}
    x_{\text{QL}}(x,z,t):= x + \chi = x+\dfrac{\delta \Lambda}{f^*}\sin(f^*t+\phi)(z-H),
\end{equation}
such that the velocity used to calculate the quasi-Lagrangian power spectrum is given by $\vec{u}'_{\text{QL}}=\vec{u}'(x_{\text{QL}}(x,z,t),z,t)$. The kinetic energy frequency spectrum indicates a clear peak at $f^*/2$, due to the subharmonic waves (Fig. \ref{fig:S1Riinv095delta08fouriertransform}, consistent with the expectation for PSI, Section \ref{section:backgroundflow}).
The subharmonic waves are localized in the BBL as there is no inertial wave imposed in the interior, and the subharmonic waves are not able to propagate into the interior for any case with $S_\infty<\sqrt{3}$ (where $f^*/2 < f/2$). 

\begin{figure}[t!]
    \centering
    \includegraphics[width=0.95\linewidth]{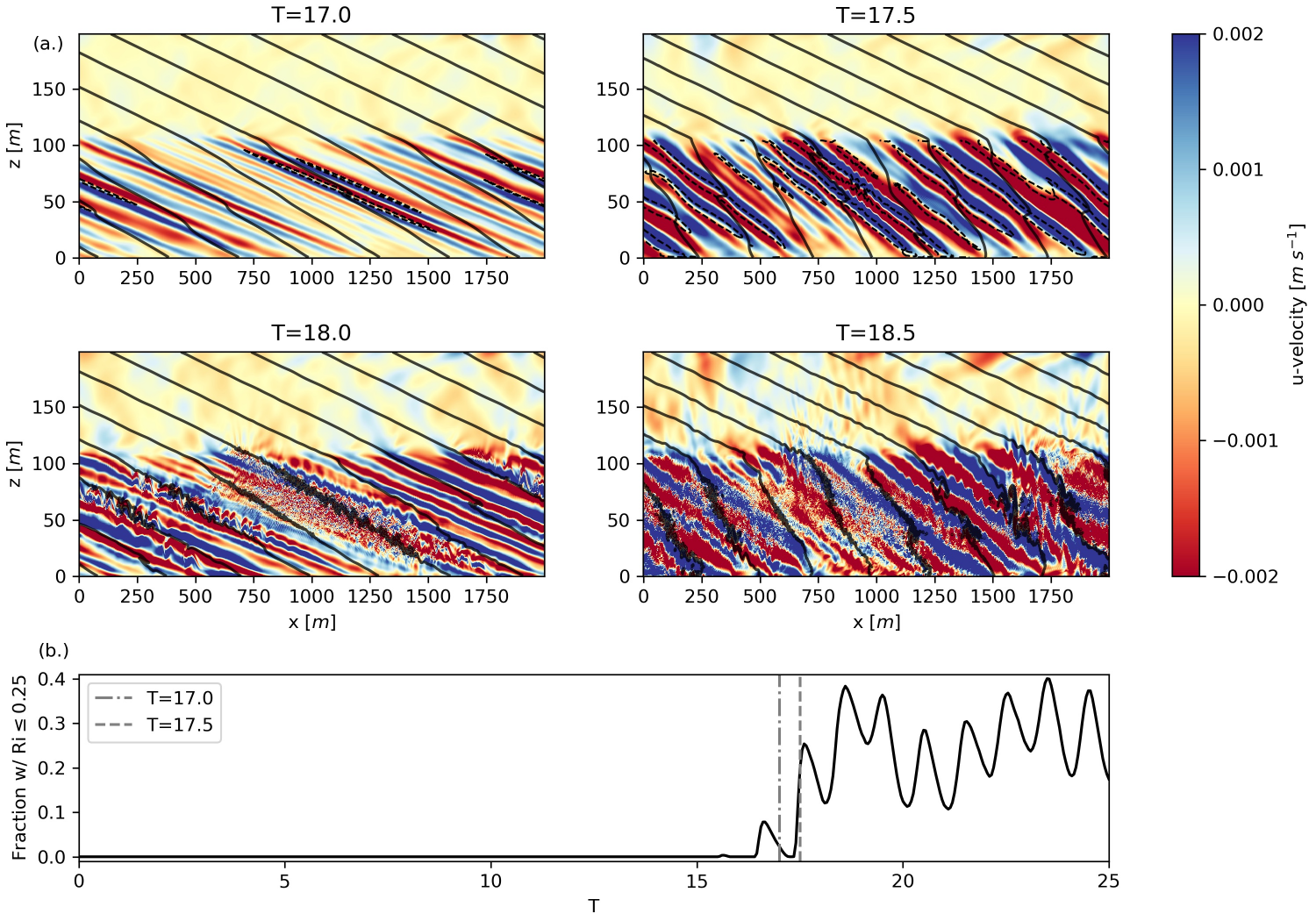}
    \caption{(a.) The $u$ perturbation velocities for Sim9 as the flow reaches finite amplitude, with isopycnals (black contours). $T:=f^*t/(2\pi)$ denotes time normalized by the modified inertial period. The black dashed contours denote regions where $\text{Ri}\leq 0.25$, and are only shown for times $T:=17.0,17.5$ as later times become noisy when the region is highlighted and it is difficult to discern the flow. (b.) The fraction of gridpoints in the BBL with a Richardson number less than $0.25$ is shown, and the points at which $T=17.0,17.5$ are marked.}
    \label{fig:S1Rinv095delta08finiteamplitude}
\end{figure}

In the BBL the instability grows until the subharmonic waves reach finite amplitude, where the instability amplitude approaches the magnitude of the imposed background flow. Sim9 experiences secondary instabilities as it reaches finite amplitude as seen in Figure \ref{fig:S1Rinv095delta08finiteamplitude}a. At $T=17.0,17.5$ the flow is reaching finite amplitude, and billows in the flow velocity and isopycnals are visible in regions where the Richardson number is less than 0.25, indicating shear instabilities. By $T=18.0,18.5$, an increase in small-scales is qualitatively evident throughout the boundary layer, and a significant portion of the domain has $Ri\le0.25$, consistent with further development of secondary shear instabilities. The 2-dimensional domain employed for these simulations prevents the development of 3-dimensional turbulence, however these results suggest that PSI will facilitate a transition to turbulence that will enhance mixing in the ocean BBL.

\begin{figure}
    \centering
    \includegraphics[width=0.95\linewidth]{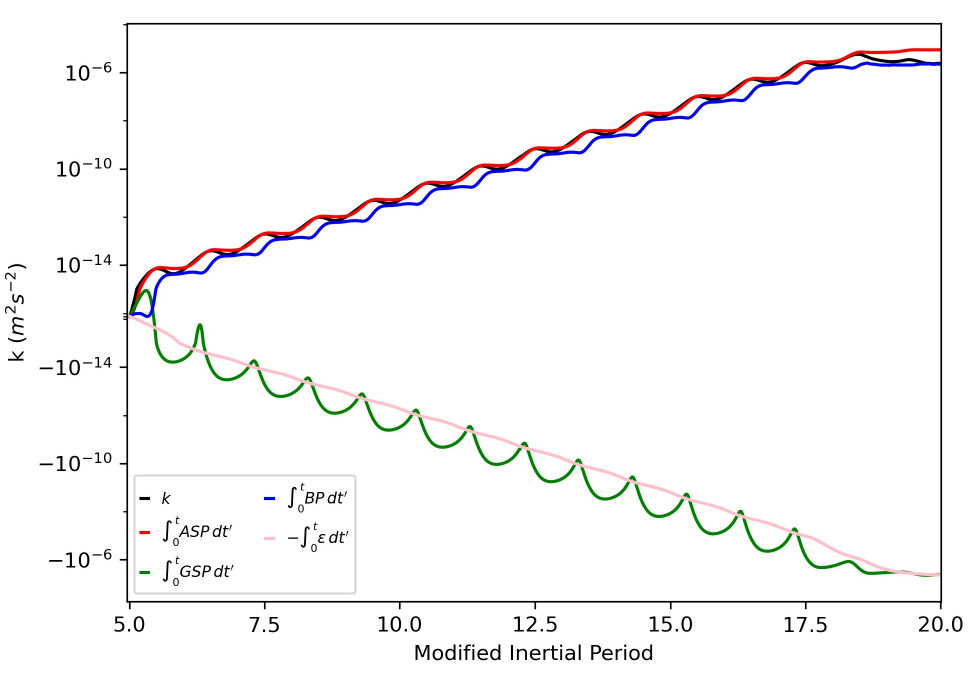}
    \caption{Time integrated TKE budget terms for Sim9. The linear threshold for the symmetric log scale is $10^{-15}$ m$^2$s$^{-2}$ and time is normalized as $f^*t/(2\pi)$.}
    \label{fig:S1Riinv095delta08integratedtke}
\end{figure}

The time and volume integrated TKE budget terms for Sim9 are shown in figure \ref{fig:S1Riinv095delta08integratedtke}. 
As expected, ASP drives growth, with positive contributions from BP and a partially compensating sink of TKE from GSP. This is consistent with previous studies on PSI in surface baroclinic flows \citep{ThomasTaylor14, Hilditch_Thomas_2023} and as was predicted by the linear stability analysis (see Figs. \ref{fig:stblty_TKE_example} and  \ref{fig:TKE_rel_contr}). Later in the simulation ASP asymptotes, reflecting the decay of the initial near-inertial oscillations due to PSI. DISS removes energy consistently throughout Sim9, with a magnitude that is typically close to GSP with less phase dependence than the production terms. 

\begin{figure}[t!]
    \centering
    \includegraphics[width=0.95\linewidth]{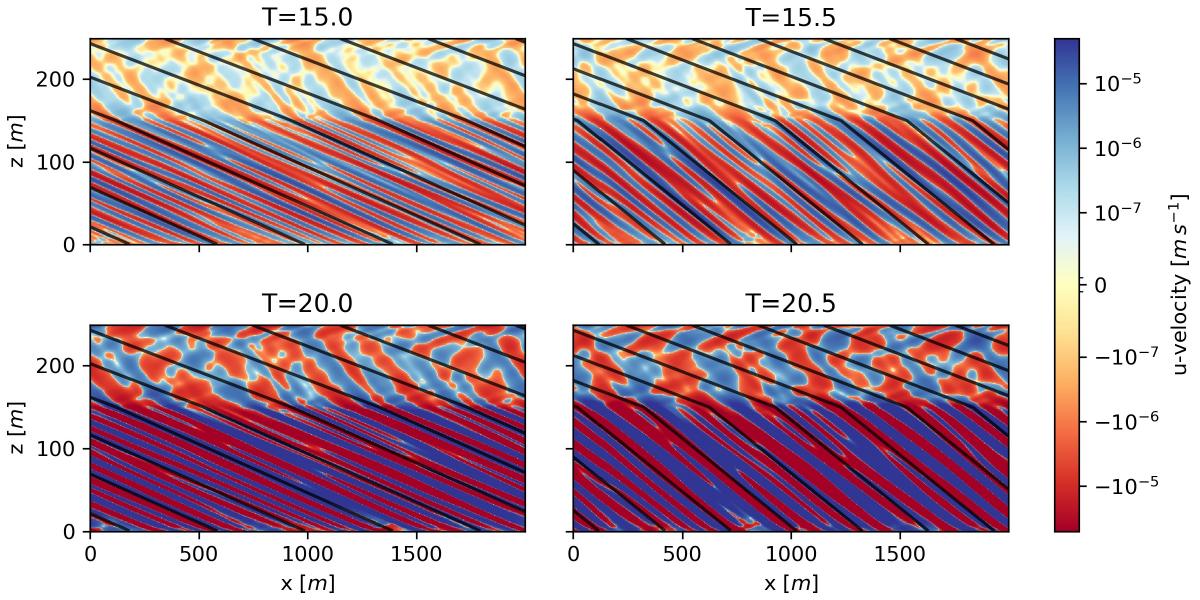}
    \caption{The $u$ perturbation velocities over the linear growth phase of PSI for Sim7, with isopcynals shown in black. The color bar is in a symmetric logarithm scale, with a linear threshold at $10^{-7}$ ms$^{-2}$. Time is normalized by the modified inertial period.}
    \label{fig:S1Rinv075delta08linearphase}
\end{figure}

\begin{figure}
    \centering
    \includegraphics[width=0.95\linewidth]{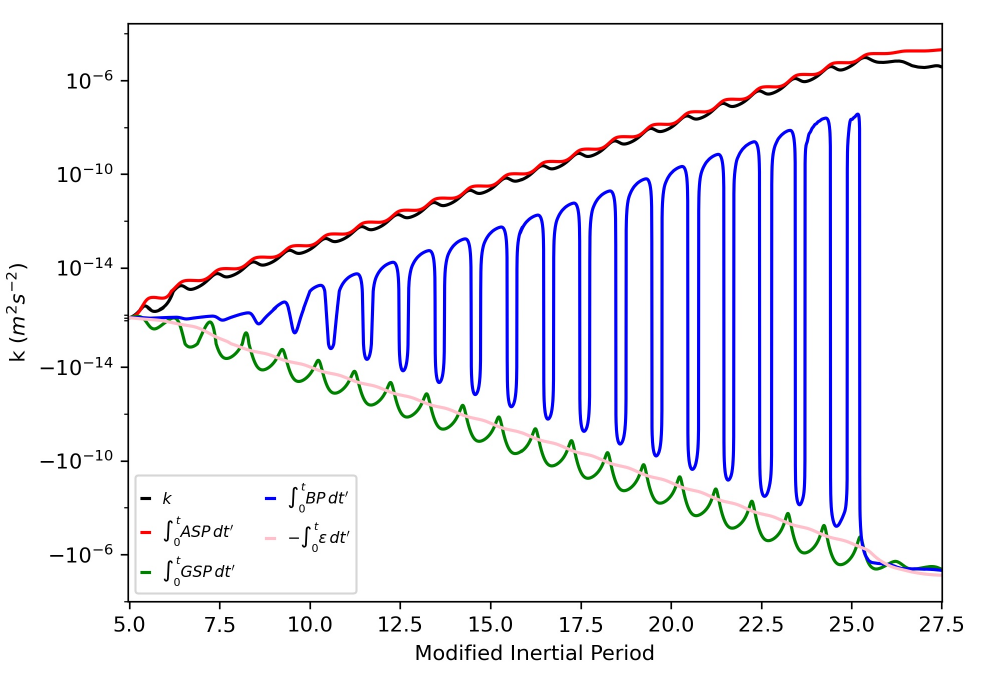}
    \caption{Time integrated TKE budget terms for Sim7. The linear threshold for the symmetric log scale is $10^{-15}$ m$^2$s$^{-2}$ and time is normalized by the modified inertial period ($f^*t/(2\pi)$).}
    \label{fig:S1Riinv075delta08integratedtke}
\end{figure}

The above progression is representative of most simulations, however as a point of comparison the results from Sim7 will also be briefly reviewed here. The perturbation slope in Sim7 more closely aligns with the the isopycnals (Fig. \ref{fig:S1Rinv075delta08linearphase}), however the structure of the subharmonic waves is somewhat less coherent. BP notably does not significantly contribute to growth in Sim7, as is seen in Figure \ref{fig:S1Riinv075delta08integratedtke}, which aligns with the predicted relative contributions of BP from the linear stability analysis. In both cases the growth of BP stalls upon reaching finite amplitude, but it is worth noting that in Sim7 the finite amplitude BP is negative whereas in Sim9 BP is positive. This behavior is not captured by the linear stability analysis (Fig. \ref{fig:BP_GSP_rel_contr}).


\section{Conclusions}\label{section:conclusions}
The potential for PSI of near-inertial waves in the ocean bottom boundary layer was demonstrated for an idealized downwelling arrested Ekman solution. The instability growth rate is determined by the strength of the wave relative to the steady thermal wind shear, $\delta$, and the ratio of the minimum allowable wave frequency to the near-inertial wave frequency, $\omega_{min}/f^*$. This latter quantity can be posed for the flow used here as a function of the interior slope Burger number, $S_\infty$, and boundary layer stratification parameter, $\gamma$, both key parameters in determining the dynamical evolution of the BBL \citep{Allen_Newberger_1998, Wenegrat_et_al_2018, Wenegrat_et_al_2020}. Floquet analysis demonstrates that PSI extends the region of instability beyond that determined by SI (Fig. \ref{fig:instability_region}), with typical growth rates of $\mathcal{O}(0.1f)$. These results broadly parallel those for surface fronts \citep{Hilditch_Thomas_2023}, with an additional dependence on $S_\infty$ that arises due to the effect of buoyancy forces in constraining across-slope motions in the BBL over sloping topography. Non-linear simulations exhibit PSI consistent with expectations from the linear stability analysis. When the simulations reach finite amplitude the flow begins to experience secondary shear instabilities that would facilitate a transition to turbulence (not explored here), providing a pathway for PSI to flux energy from internal waves to turbulence along the ocean bottom.


Turbulence near the sloping sea floor is understood to play an important role in the abyssal circulation, allowing the return of dense deep waters towards the surface \citep{Polzin1997,Ferrari_et_al_2016}. 
The canonical understanding of the mechanisms driving this mixing focuses on the role of internal waves, albeit often without consideration of the baroclinic structure of the near-bottom layer. This baroclinicity can be associated with a reduction of near-boundary PV, particularly when the interior flow is in the direction of Kelvin-wave propagation such that there is a frictional destruction of PV at the boundary \citep{Benthyusen_and_Thomas_2012}. Conditions of reduced PV over sloping topography have been observed in the key deep boundary current pathways of the overturning circulation \citep{STAHR1999, Naveira_Garabato_et_al_2019}, and recent work suggests that similar conditions may be ubiquitous along the ocean's margins \citep{schubert_ocean_2025}. These conditions have been explored previously for their role in permitting submesoscale symmetric and baroclinic instability, and the work presented here suggests the additional possibility of PSI.

The approach taken in this work assumes a steady arrested Ekman layer solution for the background flow, an idealization taken to isolate PSI from the time-dependent boundary layer adjustment. The full BBL adjustment process to downslope Ekman transport first involves the emergence of SI which acts over inertial timescales to bring the boundary layer PV back to 0 \citep{Wenegrat_et_al_2020}. At this point the flow may become unstable to PSI (given the presence of a near-inertial wave), although it is not clear how pre-existing finite-amplitude SI cells or other ageostrophic flow will alter the instability development. Likewise, in the 3-dimensional setting baroclinic modes can also develop in the BBL \citep{Wenegrat_et_al_2018}, although not permitted in the 2-dimensional configuration used here. PSI and baroclinic instability both have their fastest growth rates for small slope Burger numbers, making their joint evolution hard to predict a priori. Numerical results though suggest baroclinic modes may be strongly suppressed for steeper slopes or more stratified conditions, whereas PSI may still persist. 

Finally, here we have considered only the case of slope-parallel waves at the modified near-inertial frequency $f^*$. This is the resonant frequency of the baroclinic BBL over sloping topography \citep{Brink_Lintz_2010a, Brink_Lintz_2010b}, hence waves at this frequency are likely to be common features of the near-bottom flow, generated as a response to temporal variability of the flow or spatial variations in the bathymetry. Indeed recent moored observations of an abyssal seamount find evidence of waves at frequency $f^*$, with an estimated half of the wave energy radiated upwards along the topography and half dissipated locally \citep{Xieetal2023}. While it was not possible to directly estimate PV from these observations, the anticyclonic flow with $Ri \sim \mathcal{O}(1)$ suggest the possibility of a flow unstable to PSI. More generally however it is also of interest to understand the potential of PSI of remotely generated internal waves propagating into the baroclinic bottom boundary layer. Significant energy exists in both the baroclinic tide and in near-inertial waves generated at the surface, such that their breakdown to turbulence via PSI would provide a potentially significant source of near-bottom mixing. These waves however will often have a vertical scale much larger than the oceanic BBL, such that PSI may not occur. Future work could consider 
the conditions under which PSI of waves at other frequencies might occur in the BBL, as well as the interaction between PSI and the time-dependent boundary layer adjustment as discussed above.\\

\backsection[Acknowledgements]{We would like to acknowledge computing support from the Casper system \\(https://ncar.pub/casper) provided by the NSF National Center for Atmospheric Research (NCAR), sponsored by the National Science Foundation.}

\backsection[Funding]{L.P.K. and J.O.W gratefully acknowledge funding from the U.S. NSF grant OCE-2342990. J.O.W. was additionally supported by NSF grant OCE-2232441. L.N.T. and J.P.H. were supported by NSF grant OCE-2342991.}
\backsection[Declaration of interests]{The authors report no conflict of interest.}
%
%
%

\bibliographystyle{jfm}
\bibliography{jfm, jacob_psi}

\end{document}